\documentstyle[twocolumn,aps,psfig]{revtex}

\begin{document}
\draft
\flushbottom
\twocolumn[
\hsize\textwidth\columnwidth\hsize\csname @twocolumnfalse\endcsname

\title{ Chronology protection in a toy surface plasmon "time machine".}
\author{Igor I. Smolyaninov}
\address{ Department of Electrical and Computer Engineering \\
University of Maryland, College Park,\\
MD 20742}
\date{\today}
\maketitle
\tightenlines
\widetext
\advance\leftskip by 57pt
\advance\rightskip by 57pt

\begin{abstract}
Recently introduced toy surface plasmon "black holes" and "wormholes" (New Journal of Physics 5, 147.1-147.8 (2003), and gr-qc/0306089) can be used to create a toy "time machine" according to a number of published designs (see for example I.D. Novikov, Sov.Phys.JETP 68, 439 (1989)). Assuming that such a toy "time machine" does not work, a general prediction can be made of strong electromagnetic field enhancement inside an arbitrarily-shaped nanohole near an arising effective event horizon, which is supposed to prevent the toy "time machine" from being operational. This general result is useful in description of the nonlinear optical behavior of random nanoholes in metal films. 

\end{abstract}

]
\narrowtext

\tightenlines

Solid state toy models are useful in understanding of many complicated field-theoretical situations \cite{1}. In particular, they help in consideration of electromagnetic phenomena in curved space-time. In the case of media electrodynamics this is possible because of an analogy between the propagation of light in matter and in curved space-times: it is well known that the Maxwell equations in a general curved space-time background $g_{ik}(x,t)$ are equivalent to the phenomenological Maxwell equations in the presence of a matter background with nontrivial electric and magnetic permeability tensors $\epsilon _{ij}(x,t)$ and $\mu _{ij}(x,t)$ \cite{2}. In this analogy, the event horizon corresponds to a surface of singular electric and magnetic permeabilities, so that the speed of light goes to zero, and light is "frozen" near such a surface. Unfortunately, until recently all the suggested electromagnetic black hole toy-models were very difficult to realize and study experimentally, so virtually no experimental work was done in this field. 

Very recently a number of simple surface plasmon toy black hole and wormhole models have been suggested and observed in the experiment \cite{3,4,5}. 
These models are based on the surface plasmons that propagate along a metal surface, which is curved and/or covered with a layer of dielectric. 
Surface plasmons are collective excitations of the conductive electrons and the electromagnetic field \cite{6}. They exist in "curved three-dimensional space-times" defined by the shape of the metal-dielectric interface. Since in many experimental geometries surface plasmons are weakly coupled to the outside world (to free-space photons, phonons, single-electron excitations, etc.) it is reasonable to treat the physics of surface plasmons separately from the rest of the surface and bulk excitations, so that a field-theory of surface plasmons in a curved space-time background may be considered. For example, a nanohole in a thin metal membrane may be treated as a "wormhole" connecting two "flat" surface plasmon worlds located on the opposite interfaces of the membrane \cite{3}. On the other hand, near the plasmon resonance (which is defined by the condition that $\epsilon _m(\omega )=-\epsilon _d$ at the metal-dielectric interface, where $\epsilon _m(\omega )$ and $\epsilon _d$ are the dielectric constants of metal and dielectric, respectively \cite{6}) the surface plasmon velocity vanishes, so that the surface plasmon "stops" on the metal surface, and the surface charge and the normal component of the electric field diverge. For a given frequency of light, the spatial boundary of the plasmon resonance ("the event horizon" of our toy model) may be defined at will using the geometry $\epsilon _d(x,y)$ of the absorbed layer (or a liquid droplet) of dielectric on the metal surface. Thus, the plasmon resonance becomes a natural candidate to emulate the event horizon of a black hole. As a result, toy two-dimensional surface plasmon black holes can be easily produced and studied (see Fig.1) \cite{4,5}. In another example, a surface plasmon toy model of a rotating black hole (Kerr metric \cite{7}) can be emulated if the dielectric layer exhibits optical activity \cite{4}. 

Real wormholes and rotating black holes are the basic elements which may allow, in principle, a creation of a time machine according to a number of published designs \cite{8,9}. In what follows I am going to emulate the time machine design published in \cite{9} with toy surface plasmon "black holes" and "wormholes". Assuming that such a toy "time machine" does not work, a general prediction can be made on strong electromagnetic field enhancement inside an arbitrary-shaped nanohole near an arising effective event horizon, which is supposed to prevent the surface plasmon "time machine" from being operational. This general result is useful in description of the nonlinear optical behavior of random and artificial nanoholes in metal films, which indeed show signs of strongly nonlinear behavior in recently observed single-photon tunneling and optically-controlled photon tunneling experiments \cite{10}.

Let us recall the basic principles of a "real" time machine operation. In a time machine design of Morris, Thorne and Yurtsever \cite{8} (Fig.2(a)) at the initial time the openings of the wormhole are not far apart. Assuming that one of the openings is forced to move away at high speed (while internal distance between the openings remains unchanged and short at all times), and then to reverse its motion and return back to the vicinity of the second opening, upon completion of the motion the clock of the moving opening will lag the clock of the stationary wormhole opening. As a result, an observer passing through the wormhole may travel to his past or future depending on the direction of travel through the wormhole. Another variant of the time machine introduced by Novikov \cite{9} (Fig.2(b)) consists of a wormhole in which one opening rotates around another. The clock of a moving opening lags the clock of a stationary one, so that after some time a Cauchy horizon appears, and travel into the past becomes possible. 

Before we start to emulate one of these time machine designs with surface plasmon optics, let us recall the effective metric and the basic properties of surface plasmon toy black holes. As a first step, let us consider in detail the dispersion law of a surface plasmon (SP), which propagates along the flat metal-dielectric interface. The SP field decays exponentially both inside the metal and the dielectric. Inside the dielectric the decay exponent is roughly equal to the SP wave vector. As a first step let us assume that both metal and dielectric completely fill the respective $z<0$ and $z>0$ half-spaces. In such a case the dispersion law can be written as \cite{6} 

\begin{equation}  
k^2=\frac{\omega ^2}{c^2}\frac{\epsilon _d\epsilon _m(\omega )}{\epsilon _d+\epsilon _m(\omega)} ,
\end{equation}

where we will assume that $\epsilon _m=1-\omega _p^2/\omega ^2$ according to the Drude model, and $\omega _p$ is the plasma frequency of the metal. This dispersion law is shown in Fig.3(a) for the cases of metal-vacuum and metal-dielectric interfaces. It starts as a "light line" in the respective dielectric at low frequencies and approaches asymptotically $\omega =\omega _p/(1+\epsilon _d)^{1/2}$ at very large wave vectors. The latter frequency corresponds to the so-called surface plasmon resonance. Under the surface plasmon resonance conditions both phase and group velocity of the SPs is zero, and the surface charge and the normal component of the electric field diverge. Since at every wavevector the SP dispersion law is located to the right of the "light line", the SPs of the plane metal-dielectric interface are decoupled from the free-space photons due to the momentum conservation law. If a droplet of dielectric is placed on the metal surface, the SP dispersion law will be a function of the local thickness of the droplet. Deep inside the droplet far from its edges the SP dispersion law will look similar to the case of a metal-dielectric interface, whereas near the edges (where the dielectric is thinner) it will approach the SP dispersion law for the metal-vacuum interface. As a result, for every frequency between $\omega _p/(1+\epsilon _d)^{1/2}$ and $\omega _p/2^{1/2}$ there will be a closed linear boundary inside the droplet for which the surface plasmon resonance conditions are satisfied. Let us show that such a droplet of dielectric on the metal interface behaves as a "surface plasmon black hole" in the frequency range between $\omega _p/(1+\epsilon _d)^{1/2}$ and $\omega _p/2^{1/2}$, and that the described boundary of the surface plasmon resonance behaves as an "event horizon" of such a black hole. 

The dispersion law of surface plasmons in real experimental situations is defined by the shape and thickness of the droplet near its edge, by the thickness of the metal film, and by the dielectric properties of the substrate. In order to simplify the effective metric, let us consider a thin free-standing metal membrane with two "linear" droplets positioned symmetrically on both sides of the membrane (Fig.3(b)). The droplets thicknesses at $x=0$ correspond to the surface plasmon resonance at the illumination frequency. The droplets taper off adiabatically in positive and negative $x$-directions on both sides of the membrane. In the symmetric membrane geometry the surface plasmon spectrum consists of two branches $\omega _-$ and $\omega _+$, which exhibit positive and negative dispersion, respectively, near the surface plasmon resonance \cite{11}. Fig.3(c) shows the dispersion curves of both branches for the cases of metal-vacuum interface far from the droplets, and for the locations near $x=0$. The droplets represent an effective black hole for $\omega _-$ modes and an effective white hole for $\omega _+$ modes, similar to \cite{2}. If the effective slowly varying local surface plasmon phase velocity $c^\star (x) =\omega /k(x)$ is introduced, the wave equation for surface plasmons may be written as 

\begin{equation}
(\frac{\partial ^2}{c^{\star 2}\partial t^2}-\frac{\partial ^2}{\partial x^2}-\frac{\partial ^2}{\partial y^2})A=0,
\end{equation}

which corresponds to an effective metric

\begin{equation}
ds^2=c^{\star 2}dt^2-dx^2-dy^2
\end{equation}

The event horizon corresponds to $c^\star =0$ at $x=0$. The two regions $x>0$ and $x<0$ correspond to the two sides of a black hole, which are connected by an Einstein-Rosen bridge at $x=0$. The behavior of $c^\star (x)$ near $x=0$ may be defined at will by choosing the corresponding geometry of the droplet edge (if necessary, the droplet may be replaced by a similar shaped layer of solid dielectric). Let us assume that $c^\star =\alpha xc$ in the vicinity of $x=0$. However, we should remember that this linear behavior will be cut off somewhere near the effective horizon due to such effects as Landau damping, losses in the metal and the dielectric, etc. \cite{12}. The resulting effective metric now looks like

\begin{equation}
ds^2=\alpha ^2x^2c^2dt^2-dx^2-dy^2
\end{equation}

Thus, this particular choice of the shape of the droplet edge gives rise to an effective Rindler geometry. Similar metric would describe the space-time geometry near the event horizon of a black hole with mass $M_{BH}=c^2/8\gamma \alpha $, where $\gamma $ is the gravitation constant \cite{2}. 

Let us discuss the properties of the local "proper" time $\tau = c^\star t$, which appears in this model. This "proper" plasmon time at a metal-dielectric interface (for example, deep inside a liquid droplet) always lags the "proper" time at a metal-vacuum interface, due to the slower surface plasmon phase velocity. As a result, addition of a dielectric near metal surface emulates the slowing down of the clocks due to the motion of a reference frame or due to the gravitation field. Thus, we can try and emulate the motion of the wormhole openings in one of the time machine designs shown in Fig.2 by placing a droplet of dielectric on top of our toy wormhole (which means placing a liquid droplet on top of a hole drilled in a metal membrane (Fig.4a)). A plasmon trajectory starting from the top side of the membrane near the droplet edge and bringing it to its bottom side, and back to the top through the hole would emulate a time traveler motion described in \cite{8,9}. However, even though we would be using a distorted model "proper" time $\tau = c^\star t$ in this experiment, and there is no actual slowing down of a real clock of an observer performing experiments with surface plasmons, we are not supposed to build a surface plasmon toy "time machine": the surface plasmons cannot go back in time even according to the readings of a clock measuring this distorted "proper" time. This is clear from the fact that $c^\star $ is always positive. 

After a careful consideration we notice though, that the nature prevents operation of such a time machine: inside the hole drilled in the metal membrane another effective horizon is bound to appear. In the case of a smooth cylindrical hole partially filled with the dielectric (Fig.4b), this is quite clear from the consideration of dispersion of cylindrical surface plasmons \cite{13}. At large wavevectors ($k>>1/R$, where R is the cylinder radius) the dispersion law of cylindrical plasmons looks similar to the dispersion of regular surface plasmons (eq.(1)). Thus, virtually the same consideration as above may be repeated for cylindrical plasmons near the meniscus boundary, and we come to a conclusion that an effective horizon should appear near the meniscus boundary. 

However, the impossibility of building a time machine implies a much stronger conclusion, which would be valid in any arbitrarily-shaped pinhole in a metal film covered with a dielectric: an effective horizon must appear in any such pinhole. The value of this conclusion is in its generality. Maxwell equations are not easy to solve for some random, asymmetric, arbitrarily-shaped pinholes. The consideration above predicts strong electromagnetic field enhancement in any such geometry, which means that nonlinear optical properties of pinholes covered with dielectrics should be strongly enhanced. This conclusion agrees well with our experimental observations \cite{10}.

Nonlinear optical effects put limits on the validity of the effective metrics (3) or (4). While, this statement is true in general, it may not be valid for the lowest order nonlinearities. For example, the nonlinearities of the liquid dielectric of the form

\begin{equation}
n=n_0+n_2E^2 ,
\end{equation}

where $E$ is the local electric field and $n_2>0$, which would be responsible for the self-focusing effect in three-dimensional optics, may lead to a toy "gravitational collapse" of the surface plasmon field near the toy "black hole". This type of nonlinearity causes an effective gravitational interaction of surface plasmons with each other. Thus, we may imagine a situation where a liquid droplet is illuminated with an intense plasmon beam at a frequency below $\omega _p/(1+\epsilon _d)^{1/2}$, so that a low intensity plasmon field would not experience an event horizon near the droplet edge. However, the increase in the droplet refractive index due to the high intensity plasmon field will cause the plasmon field to collapse towards an arising event horizon. Such a self-focusing effect may cause even stronger local field enhancement in the droplet and nanohole geometries observed in our experiments \cite{10}. 

In conclusion, recently introduced toy surface plasmon "black holes" and "wormholes" can be used to create a toy "time machine" according to a number of published designs \cite{8,9}. Assuming that such a toy "time machine" does not work, a general prediction can be made of strong electromagnetic field enhancement inside an arbitrarily-shaped nanohole near an arising effective event horizon, which is supposed to prevent the toy "time machine" from being operational. This general result is useful in description of the nonlinear optical behavior of random nanoholes in metal films.

This work has been supported in part by the NSF grants ECS-0210438 and ECS-0304046.

Figure captions.

Fig.1 Far-field images of a toy surface plasmon black hole: (a) Droplet of glycerin on a gold film surface (illuminated from the top). The droplet diameter is approximately 15 micrometers. (b) The same droplet illuminated with white light in the Kretschman geometry, which provides efficient coupling of light to surface plasmons on the gold-vacuum interface. The white rim around the droplet boundary corresponds to the effective surface plasmon "event horizon". 

Fig.2 (a) Time machine design of Morris, Thorne and Yurtsever. Wormhole opening A is stationary, while the opening B moves along the Z-direction and returns back. (b) Time machine design of Novikov. Wormhole opening A is stationary, while the opening B rotates around it. 

Fig.3 (a) Surface plasmon dispersion law for the cases of metal-vacuum and metal-dielectric interfaces. (b) A thin metal membrane with two "linear" droplets positioned symmetrically on both sides of the membrane. An effective event horizon is located at $x=0$. (c) Dispersion laws of the $\omega _-$ and $\omega _+$ modes exhibit positive and negative dispersion, respectively, near the surface plasmon resonance. These branches are shown for the cases of metal-vacuum interface far from the droplets, and for the locations near $x=0$.  

Fig.4 (a) Surface plasmon toy "time machine": a droplet of dielectric is placed on top of the hole drilled in a metal membrane near its opening B, in order to emulate its motion. The metal membrane is made thicker to the right of the hole in order to allow plasmon propagation from the top side of the membrane to its bottom side, and back to the top through the hole (opening A). (b) Cylindrical plasmons inside a smooth cylindrical channel partially filled with dielectric experience an effective event horizon at the edge of the meniscus. 


\begin{references}

\bibitem{1} G.E. Volovik, The Universe in a Helium Droplet (Clarendon, Oxford, 2003) and the references therein.

\bibitem{2} B. Reznik, Phys.Rev.D 62, 044044 (2000).

\bibitem{3} I.I. Smolyaninov, Phys.Rev.B 67, 165406 (2003). 

\bibitem{4} I.I. Smolyaninov, New Journal of Physics 5, 147.1-147.8 (2003); I.I. Smolyaninov and C.C. Davis, gr-qc/0306089.

\bibitem{5} I.I. Smolyaninov, cond-mat/0309590.

\bibitem{6} H. Raether, Surface Plasmons, Springer Tracts in Modern Physics Vol.111 (Springer, Berlin, 1988).

\bibitem{7} L.D. Landau and E.M. Lifshitz, Field Theory (Pergamon, New York, 1984).

\bibitem{8} M.S. Morris, K.S. Thorne, and U. Yurtsever, Caltech Preprint GRP-164 (1988); M.S. Morris and K.S. Thorne, Caltech Preprint GRP-067 (1987).

\bibitem{9} I.D. Novikov, Sov.Phys.JETP 68, 439 (1989). 

\bibitem{10} I.I. Smolyaninov $et$ $al.$, Phys.Rev.Lett. 88, 187402 (2002); I.I. Smolyaninov $et$ $al.$, Phys.Rev.B 66, 205414 (2002).

\bibitem{11} K.L. Kliewer and R. Fuchs, Phys.Rev. 144, 495 (1966).

\bibitem{12} A.G. Malshukov, Phys.Reports 194, 343 (1990).

\bibitem{13} U. Schroter and A. Dereux, Phys.Rev.B 64, 125420 (2001).


\end{references}
\end{document}